\documentclass[12pt]{article}
\usepackage[left=1.25in,top=1in]{geometry}
\usepackage{amsthm}
\usepackage{amsmath}
\usepackage{graphicx,url}
\usepackage{booktabs}
\usepackage[small,bf]{caption}

\newcommand{\Oh}[1]
  {\ensuremath{\mathcal{O}\!\left({#1}\right)}}
\newcommand{\access}
  {\ensuremath{\mathrm{access}}}
\newcommand{\rank}
  {\ensuremath{\mathrm{rank}}}
\newcommand{\select}
  {\ensuremath{\mathrm{select}}}

\newtheorem{obs}{Observation}

\newtheorem{thrm}[obs]{Theorem}
\newtheorem{cor}[obs]{Corollary}

\begin{document}

\title{\vspace{-4ex}Queries on LZ-Bounded Encodings\thanks{This work is supported by the Academy of Finland.}}
%\title{Rank and Select in Lempel-Ziv Compressed Strings\thanks{This work is supported by the Academy of Finland.}}

\author{\normalsize Djamal Belazzougui$^1$, Travis Gagie$^1$, Pawe\l\ Gawrychowski$^2$,\\ 
\normalsize Juha K\"arkk\"ainen$^1$, Alberto Ord\'o\~{n}ez$^3$, Simon J.\ Puglisi$^1$, and Yasuo Tabei$^4$\\[1ex]
\footnotesize $^1$ Helsinki Institute for Information Technology (HIIT) and\\[-0.8ex]
\footnotesize Department of Computer Science, University of Helsinki, Finland\\[-0.8ex]
\footnotesize $^2$ Max Planck Institute for Informatics, Germany\\[-0.8ex]
\footnotesize $^3$ Database Lab, University of A Coru\~{n}a, Spain\\[-0.8ex]
\footnotesize $^4$ PRESTO, Japan Science and Technology Agency, Japan}
\date{}
\maketitle
\vspace{-4ex}

\begin{abstract}
We describe a data structure that stores a string $S$ in space similar to that of its Lempel-Ziv encoding and efficiently supports access, rank and select queries. These queries are fundamental for implementing succinct and compressed data structures, such as compressed trees and graphs. We show that our data structure can be built in a scalable manner and is both small and fast in practice compared to other data structures supporting such queries.
\end{abstract}

\section{Introduction}
\label{sec:introduction}

A common approach in the design of compressed data structures is to translate operations on
the original (uncompressed) data structure into simple queries over compressed strings.
Perhaps the most fundamental of these queries are access, rank and select. Given a string $S$
of symbols drawn from an alphabet of size $\sigma$, these queries are defined as:
\begin{align*}
  \access(i,j) &= \text{return the substring $S[i,j]$} \\
  \rank_a(i) &= \text{return the number of occurrences of the character $a$}\\
  &\ \ \ \ \text{among the first $i$ symbols of $S$} \\
  \select_a(j) &= \text{return the position of the $j$th occurrence of $a$.} 
\end{align*}

There have been dozens of papers written about how to support fast access, rank and select queries on compressed strings, and even more about how to use those queries when building other compressed data structures; see, e.g.,~\cite{RR08,Gag??} for surveys.  Only a few of those papers, however --- e.g.,~\cite{NPV11,NO14,BPT14} --- have considered how to support rank and select 
on LZ77- or grammar-compressed strings.  This is an important problem when, e.g., compressing rooted, unlabelled trees with 
many repeated subtrees (such as the shapes of suffix trees~\cite{NOsea14.1} or XML parse trees~\cite{LMM13}) while still supporting 
fast navigation in them.

In this paper show how to adapt block graphs~\cite{GGP11,GHP14}, originally designed to support only access, so that they support fast rank and select queries as well, with various time-space tradeoffs.
We note that the block graphs we use here are trees, while those in~\cite{GGP11,GHP14} were directed 
acyclic graphs but not trees (which is why they are not called ``block trees''). 
We also note that block graphs are collage systems~\cite{KMSTSA03} but even now they are still not context-free grammars.

Our main results are as follows:  We can store a string \(S [1..n]\) over an alphabet of size $\sigma$ in $\Oh{zr \log (n) \log_r \frac{n\log\sigma}{z\log n}}$ bits, where $z$ is the number of phrases in the LZ77 parse of $S$ and \(r \leq n\), such that we can support extraction of a substring of length $m$ in \(\Oh{\log_r \left( \frac{n\log\sigma}{z\log n} \right) \cdot \left( \frac{m \log \sigma}{\log n} +1 \right)}\) time.  Using a $\sigma$ factor more space, we can support rank in $\Oh{\log_r \frac{n\log\sigma}{z\log n}}$ time and select in $\Oh{\log_r \left( \frac{n\log\sigma}{z\log n} \right) \log \log n}$ time.

\section{Block Graphs}
\label{sec:block_graphs}

If \(n < r\) then the block graph of degree $r$ for \(S [1..n]\) is a single node that stores $S$.  If \(n \geq r\) then the root of the block graph has $r$ children.  To determine which of these children are leaves and which are internal nodes, we divide $S$ into blocks \(S_1, \ldots, S_r\) such that \(|S_1| = \cdots = |S_{n \bmod r}| = \lceil n / r \rceil\) and \(|S_{(n \bmod r) + 1}| = \cdots = |S_r| = \lfloor n / r \rfloor\).  For \(1 \leq i < r\), if \(S_i S_{i + 1}\) is the leftmost occurrence in $S$ of that substring --- which is the case if \(S_i S_{i + 1}\) contains the leftmost occurrence in $S$ of {\em any} substring --- then we mark both $S_i$ and $S_{i + 1}$.

If $S_i$ is unmarked, then the root's $i$th child is a leaf that stores: 1) pointers to one or two of its left siblings --- which must be marked --- whose corresponding blocks contain the leftmost occurrence of $S_i$, and 2) the offset of that occurrence in those blocks.  If $S_i$ is marked and \(|S_i| \leq r\) then the leaf instead stores $S_i$.  Otherwise, the child is an internal node with $r$ children.  In the latter case, we divide $S_i$ into $r$ sub-blocks as evenly as possible such that larger sub-blocks precede smaller ones.  If a consecutive pair of sub-blocks are the leftmost occurrence in $S$ of that substring --- and, thus, contained in some marked block or consecutive pair of marked blocks --- then we mark both those sub-blocks.

If the $j$th sub-block of a block has length greater than $r$ but is unmarked, then the child's $j$th child is a leaf storing pointers to its one or two left siblings whose corresponding sub-blocks contain the leftmost occurrence of the $j$th sub-block, and the offset of that occurrence in those sub-blocks.  If the sub-block is marked and has length at most $r$, then the leaf instead stores that sub-block.  Otherwise, the child's child itself has $r$ children.  Continuing this recursion, we eventually obtain a $r$-ary tree of height $\log_r n$.

\begin{figure}[t]
\includegraphics[width=\textwidth]{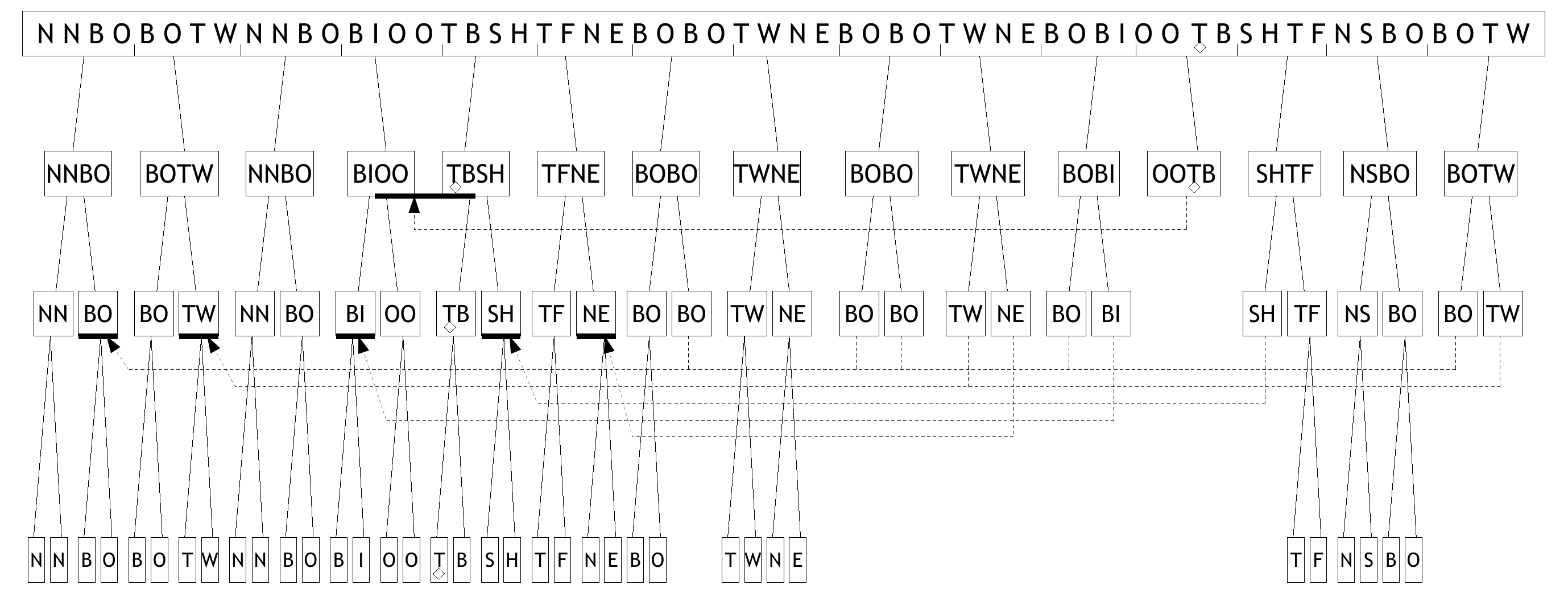}
\caption{A block graph for {\tt NNBOBOTW\dots NSBOBOTW} (from ``99 Bottles of Beer on the Wall''): black boxes are nodes; solid lines are edges and dashed ones are pointers to blocks' first occurrences; only the symbols at the leaves are stored; the `{\sf T}' symbols decorated with a $\diamond$ illustrate an access.  Notice that, although {\tt BISH} occurs only once, we do not mark the 22nd and 23rd blocks at the third level, since they are not consecutive in the original string.}
\label{fig:graph}
\end{figure}

We can stop the recursion when the cost of storing a block becomes less than that of storing a pointer --- i.e., when the blocks have size $\Oh{\log (n) / \log \sigma}$ --- which reduces the height to $\log_r \frac{n \log \sigma}{\log n}$.  If we know $z$, then we can further reduce the height to $\log_r \frac{n \log \sigma}{z \log n}$ by dividing $S$ into \(r z\) blocks but then recursing as before; this skips the first \(\log_r z\) rounds of the recursion and levels in the tree, at the cost of increasing the size by an $\Oh{z r}$ term (which will not change our asymptotic analysis).  Figure~\ref{fig:graph} shows an example of a block graph.

\section{Analysis}
\label{sec:analysis}

The first level of the block graph contains blocks of size
$n/r$, the second level blocks of size $n/r^2$, and so on, until
the last level, which has blocks of size $\log (n) / \log \sigma$. Those lowest
blocks are stored in plain form, as their original substrings.
By construction, the total number of blocks at any level never exceeds
$zr$, where $z$ is the number of phrases in the LZ77 parsing of the string.

At any level $i$ except the last, the $t_i$ blocks are encoded
using $\Oh{t_i\log n}$ bits of space, for storing pointers of $\log n$ bits
each into level $i$ data. At the last level, each block is simply encoded
as plain text, i.e., $\log (n) / \log \sigma$ symbols of size $\log\sigma$ bits each,
which is $\log n$ bits per block.

Since the upper $\log_r z$ levels contain a geometrically increasing number of blocks upper bounded by $z$,
their total encoding size is $\Oh{z\log n}$ bits. Then, each of the remaining
$\log_r \frac{n \log \sigma}{z \log n}$ levels will be encoded using $\Oh{zr \log n}$ bits
each, for a total of $\Oh{zr\log n\log_r \frac{n \log \sigma}{z \log n}}$ bits.

The query time of the block graph will be upper bounded by the number
of levels.  As mentioned in Section~\ref{sec:block_graphs}, in order to improve the time
without sacrificing our asymptotic space bound, we will start the construction of the block
graph from level $\log_r z$. Then, the number of levels is reduced to
$\log_r \frac{n \log \sigma}{z \log n}$ and the bound $\Oh{zr}$ on the number of blocks
per level still applies.

\begin{thrm}
Given a string $S$ of length $n$ over alphabet of size $\sigma$ and a parameter
$r$, we can build a block graph with $\log_r \frac{n \log \sigma}{z \log n}$ levels,
where $z$ is the number of phrases in the Lempel-Ziv parsing of $S$. The block
graph occupies a total of $\Oh{zr \log (n) \log_r \frac{n \log \sigma}{z \log n}}$ bits of space.
\end{thrm}

We note that $\frac{n \log \sigma}{z \log n}$ is actually the compression
ratio. An interesting setting for the arity is $r= \left(\frac{n \log \sigma}{z \log n}\right)^\epsilon$
for some constant $\epsilon<1$. This makes the space usage
$\Oh{\frac{(z\log n)^{1-\epsilon}(n\log\sigma)^\epsilon}{\epsilon}}$
and the number of levels $\Oh{1/\epsilon}$. This space usage is a weighted
geometric average between $z\log n$ (the space usage achievable by the
Lempel-Ziv parsing) and the original space $n\log\sigma$. The parameter
$\epsilon$ allows us to give an arbitrarily large weight to the space usage
at the cost of increasing the number of levels (and thus query time).

\section{Queries}
\label{sec:queries}

The simplest query to answer with a block graph is to return a character \(S [i]\) when given $i$.  To do this, we start at the root and descend to the child whose corresponding block contains \(S [i]\), then to the grandchild whose block contains \(S [i]\), etc.  If we reach a leaf $v$, then either $v$ stores its block explicitly --- and so we can return \(S [i]\) immediately --- or $v$ stores pointers to its left siblings whose blocks contain the leftmost occurrence in $S$ of $v$'s block, and the offset of that occurrence in those blocks.
In the latter case, in $\Oh{1}$ time we can identify a character \(S [i']\) in one of those left siblings' blocks such that \(S [i'] = S [i]\), then start descending from that left sibling to find \(S [i']\).  Returning \(S [i]\) takes a total of $\Oh{\log_r \frac{n \log \sigma}{z \log n}}$ time, proportional to the height of the tree.

For example, to return \(S [37]\) with the block graph shown in Figure~\ref{fig:graph}, we first descend to the twelfth child of the root, which is a leaf; since \(S [37]\) is the third character in that child's block \(S [35..38]\) and the first occurrence of that block is \(S [15..18]\), we know \(S [37] = S [17]\).  We follow the pointer to the root's fifth child, with block \(S [17..20]\); we then descend two more levels and eventually return {\tt T}.  The relevant characters are marked with diamonds.

\subsection{Access}
\label{subsec:access}

The original paper on block graphs~\cite{GGP11} showed how to store $S$ in $\Oh{z \log (n / z)}$ space such that any 
substring of length $m$ can be extracted in $\Oh{\log n + m}$ time. In this section we describe a better result, 
showing how to extract an arbitrary substring from a block graph with $\ell$ levels in $\Oh{\ell(m/\log_\sigma n+1)}$ time.

In order to achieve the improved bounds, we store the first and the last $\log_\sigma n$ symbols for every 
block at every level. This adds $\log n$ bits per block and does not increase the space asymptotically.
We extract a substring $S [i.i + m - 1]$ with $m \leq \log_\sigma n$ as follows.      
At the upper level, we check whether $S [i.i + m - 1]$ spans two blocks or is contained in one single block. If it spans two blocks, then we can extract the part of the string that lies in the first block in 
constant time, since we have stored the last $\log_\sigma n$ characters of the block. The same goes 
for the part that lies in the second block, since we have stored the first $\log_\sigma n$ characters 
of the block. If $S [i.i + m - 1]$ is fully contained in a block, then we descend to the next level, 
either directly if the block exists at the next level, or by following a pointer if the block was 
copied.  We continue recursively in this way, stopping either at the first level at which $S [i.i + m - 1]$ spans
two blocks, or when we reach the last level of the block graph (where all the blocks are of length $\log_\sigma n$
and their content is stored explicitly). Overall, the time spent is $\Oh{\ell}$.  To extract $S [i.i + m - 1]$ when \(m > \log_\sigma n\), we simply divide it into pieces of length \(\log_\sigma n\) (except that the last may be shorter) and extract each piece separately.

\begin{thrm}
Given a string $S$ of length $n$ over alphabet of size $\sigma$ and a parameter
$r$, we can build a data structure occupying $\Oh{zr\log_r \frac{n\log\sigma}{z\log n}}$ 
bits of space that allows extraction of any substring of $S$ of length $m$
in time
\[\Oh{\log_r \left( \frac{n\log\sigma}{z\log n} \right) \cdot \left( \frac{m \log \sigma}{\log n} +1 \right)}\,.\]
\end{thrm}

Setting $r = (n\log\sigma/(z\log n))^\epsilon$, we obtain the following corollary.

\begin{cor}
Given a string $S$ of length $n$ over alphabet $[1..\sigma]$ and a constant
$\epsilon<1$, we can build a block graph with $\Oh{1/\epsilon}$  levels.
The block graph occupies a total of $\Oh{\frac{(z\log n)^{1-\epsilon}(n\log\sigma)^\epsilon}{\epsilon}}$ bits of space,
where $z$ is the number of phrases in the LZ77 parsing of $S$,
and allows extraction of any substring of $S$ of length $m$
in time $\Oh{\lceil\frac{m}{\log_\sigma n}\rceil/\epsilon}$.
\end{cor}

\subsection{Rank}
\label{subsec:rank}

To support rank quickly on $S$, for each character $a$, we store at each node the number of occurrences of $a$ in the prefix of $S$ preceding the corresponding block.  This takes $\Oh{\sigma z r \log (n) \log_r \frac{n\log\sigma}{z\log n}}$ space.  This sample of rank values lets us turn any rank query on $S$ into a rank query on a block in $\Oh{1}$ time.  We also store information that lets us turn any rank query on an unmarked block into a rank query on a marked block in $\Oh{1}$ time.  With our sample, we can also turn a rank query on a marked block for an internal node, into a rank query on one of its children, also in $\Oh{1}$ time.  We store a rank data structure for the concatenation of the marked blocks for leaves, which takes \(\Oh{z r \log (\sigma) / \log n} = \Oh{z r}\) space, so we can answer rank queries on those blocks directly in $\Oh{1}$ time, and can thus answer rank queries on $S$ in $\Oh{\log_r \frac{n\log\sigma}{z\log n}}$ total time.

\begin{figure}[t]
\begin{center}
\includegraphics[width=.3\textwidth]{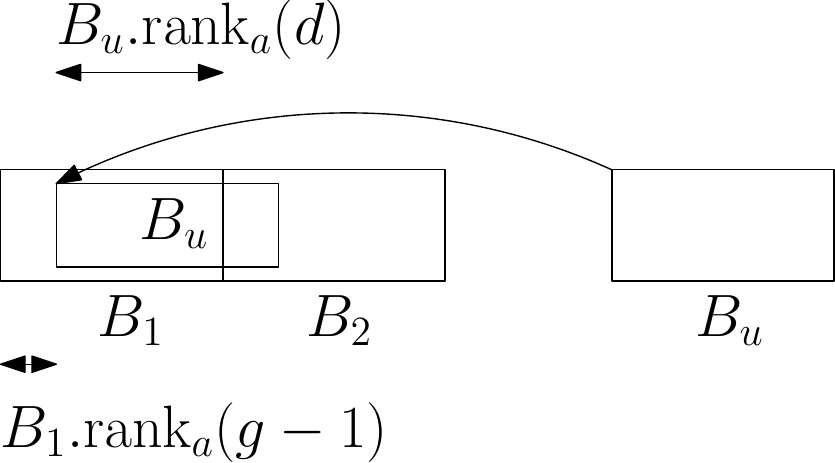}
\caption{To be able to turn a rank query on the unmarked block $B_u$ into a rank query on one of the consecutive pair of marked blocks $B_1$ and $B_2$ that contain $B_u$'s first occurrence in $S$, we store \(B_1.\rank_a (g - 1)\) and \(B_u.\rank_a (d)\).  We already have stored the offset $g$ of the occurrence of $B_u$ in \(B_1 B_2\) and we can compute from $g$ and $|B_1|$ the length $d$ of the prefix of $B_u$ that is a suffix of $B_1$.}
\label{fig:rank}
\end{center}
\end{figure}

The information that lets us change any rank query on an unmarked block $B_u$ into a rank query on a marked block is, first of all, the pointers to the marked block $B_1$ or consecutive pair $B_1$ and $B_2$ of marked blocks containing the first occurrence of $B_u$, and the offset $g$ of that occurrence in \(B_1 B_2\).  Secondly, for each character $a$, we store the number \(B_1.\rank_a (g - 1)\) of occurrences of $a$ in the prefix of $B_1$ before the occurrences of $B_u$.  Notice we can compute, from the offset $g$ and the length of $B_1$, the length $d$ of the prefix of $B_u$ that is a suffix of $B_1$.  Finally, we store the number \(B_u.\rank_a (d)\) of occurrences of $a$ in this prefix of $B_1$.  If \(i < d\) then \(B_u.\rank_a (i) = B_1.\rank_a (g + i) - B_1.\rank_a (g - 1)\).  If \(i = d\) then we have the answer stored.  If \(i > d\) then \(B_u.\rank_a (i) = B_u.\rank_a (d) + B_2.\rank_a (i - d)\).  See Figure~\ref{fig:rank}.

\subsection{Select}
\label{subsec:select}

To support select quickly on $S$, we store a predecessor data structure on the rank samples at the beginnings of the blocks.  If we use a trie with branching factor $n^{\epsilon / 2}$, we can use $\Oh{z n^\epsilon}$ space in total for the whole block graph and support predecessor queries in $\Oh{1}$ time.  On the other hand, if we use an $\Oh{z r \log_r n}$-space data structure, then predecessor queries take $\Oh{\log \log n}$ time.  This predecessor data structure lets us turn any select query on $S$ into a select query on a block, and turn any select query on a marked block for an internal node into a select query on one of its children.  The information we already have stored lets us turn any select query on an unmarked block into a select query on a marked block: if \(j \leq B_u.\rank_a (d)\) then \(B_u.\select_a (j) = B_1.\select (j + B_1.\rank_a (g - 1)) - g\).  If \(j > B_u.\rank_a (d)\) then \(B_u.\select_a (j) = B_2.\select (j - B_u.\rank_a (d)) + d\).  It follows that answering select queries on $S$ takes an $\Oh{\log_r n}$-factor more time than answering a predecessor query.

Combining the bounds for all the queries, we obtain the following result:

\begin{thrm}
\label{thm:main}
We can store a string \(S [1..n]\) over an alphabet of size $\sigma$ in
\[\Oh{zr \log (n) \log_r \frac{n\log\sigma}{z\log n}}\]
bits, where $z$ is the number of phrases in the LZ77 parse of $S$ and \(r \leq n\), such that we can support extraction of a substring of length $m$ in
\[\Oh{\log_r \left( \frac{n\log\sigma}{z\log n} \right) \cdot \left( \frac{m \log \sigma}{\log n} +1 \right)}\]
time.  Using a $\sigma$ factor more space, we can support rank in $\Oh{\log_r \frac{n\log\sigma}{z\log n}}$ time and select in $\Oh{\log_r \left( \frac{n\log\sigma}{z\log n} \right) \log \log n}$ time.  In particular, if we use $\Oh{\sigma z n^\epsilon}$ space, then rank and select take constant time and extraction takes optimal $\Oh{m \log (\sigma) / \log n + 1}$ time.
\end{thrm}

\subsection{Lowest Common Ancestor}
\label{subsec:lca}

Many queries on trees involve computing nodes' lowest common ancestors.  Consider the balanced-parentheses representation of a tree and let $S$ be the binary string we obtain by replacing each opening parenthesis by a +1 and each closing parenthesis by a -1.  Finding the lowest common ancestor of two given nodes reduces to finding the position $j$ in a given range \(S [i..k]\) such that the partial sum \(S [i] + \cdots + S [j]\) is minimum; we will give more details in the full version of this paper. 

We store the information we need to compute rank quickly on $S$, which takes $\Oh{zr \log (n) \log_r \frac{n}{z\log n}}$ bits.
We also store the minimum partial sum (computed from the beginning of $S$) in each block.  Thirdly, we store a position-only range-minimum data structure over the partial sums of the concatenation of the blocks at the bottom level, which takes $\Oh{z r \log n}$ bits.  We also store, for each other level in the block graph, a position-only range-minimum data structure over the string containing the minimum partial sum (computed from the beginning of $S$) from each block at that level.  This takes $\Oh{1}$ bits per block in the block graph.

To find $j$, we divide \(S [i..k]\) into $\Oh{\log_r \frac{n}{z\log n}}$ sub-ranges, each exactly covered by a consecutive set of blocks at some level, except that the first and last sub-ranges may each have length less than \(\log n\) and be properly contained in single blocks at the bottom level.  We use our first range-minimum data structure to find the positions of the minimum partial sums in the first and last sub-ranges.  For each other level, we our range-minimum data structures for that level to find the position of the minimum partial sum in the sub-range for that level.  Notice that, since each of those sub-ranges consists of complete blocks, the position of the minimum partial sum computed from the beginning of $S$, is the same as the position of the minimum partial sum computed from the beginning of the sub-range.

This leaves us with $\Oh{\log_r \frac{n}{z\log n}}$ candidates for the position of the minimum partial sum in \(S [i..k]\).  We use rank queries to compute the partial sum $p$ of the prefix of $S$ ending at \(S [i - 1]\), and the the minimum partial partial sums (computed from the beginning $S [i]$ of the query range) in the first and last sub-ranges.  We subtract $p$ from the partial sums for the other candidate positions, so they are computed from $S [i]$ instead of the beginning of $S$, and return return the position of the candidate position with the minimum adjusted partial sum.  In total, we use $\Oh{\log_r \frac{n}{z\log n}}$ time.

\section{Construction}
\label{sec:construction}
We now describe an algorithm for block graph construction in the External Memory Model with
memory size $M$ and block transfer size $B$. The algorithm builds a block graph of size 
$\Oh{zr \log (n) \log_r \frac{n\log\sigma}{\log n}}$ bits with $\log_r \frac{n\log\sigma}{\log n}$ 
levels.\footnote{The construction algorithm we describe cannot achieve the ideal bound of $\Oh{\log_r \frac{n \log \sigma}{z\log  n}}$ levels directly because it operates without prior knowledge of $z$, but we can remove the top levels later.}.
Construction consists of two phases. The first phase constructs the block graph in 
$\log_r (n/\log_\sigma n))$ iterations and is Monte-Carlo, so may fail with very small 
probability. The second phase attempts to reconstruct the input string from the constructed 
block graph and in doing so verifies that the block graph is indeed correct.

At the first iteration of the first phase, the algorithm processes the input
into blocks of size $b_1=n/r$, at second iteration into blocks of size $b_2=n/r^2$, and so on.
Thus at iteration $i$ we have $b_i=n/r^i$.
Each iteration in the first phase, involves two scans.
In the first scan we generate the Karp-Rabin signature of each block and store it in a hash table,
along with the block's starting position.
In the second scan, we slide a window of length $b$ over the input. To process the substring in the window at some step,
we compute
its Karp-Rabin signature and inspect the hash table. In this way we are able to determine previous
occurrences of the blocks, should they exist.
We assume we have enough internal memory for the hash table and the final block graph.

Consider the cost of the first (Monte-Carlo) phase.
In the first $\log_r(z)$ iterations, the amount of data scanned will stay $\Theta(n/(B\log_\sigma n))$
and the two scans involved take $\Theta(n/(B\log_\sigma n))$ I/Os. In the subsequent
$\log_r(n/(z\log_\sigma n))$ iterations, the amount of data scanned is reduced by a factor of $r$ in
each iteration. Thus the amount of scanned data is geometrically decreasing and the total cost is dominated
by the first such iteration, which is $\Theta(n/(B\log_\sigma n))$.
The overall cost of the Monte-Carlo phase is thus
$\Oh{(n/(B\log_\sigma n)}\cdot\log_r(z/\log_\sigma n))$.

To make the algorithm Las-Vegas, we construct a text from the block graph and compare
it with the input text. If they match (which happens with high probability), we are done.
Otherwise, we rerun the Monte-Carlo phase, and repeat the verification, and so on.
This makes block graph construction time expected, but correctness certain. 
We are able to show that reconstructing a text from the block graph takes
$$\Oh{ (n/(B\log_\sigma n))\cdot \mathtt{max}\{\log_{M/B}(n/(B\log_\sigma n)),\log_r(n/(z\log_\sigma n))\}}$$
I/Os; however, due to lack of space, we defer the details to the full article. 

\section{Experiments}
\label{sec:experiments}

We have implemented our data structure and in this section we report on its
practical performance in comparison to other state-of-the art solutions. Due 
to lack of space, in this extended abstract we only provide experimental results 
for rank, select, and access operations, leaving treatment of range minimum 
and previous smaller value to the full article. Our implementation of select
diverges somewhat from the description in Section~\ref{subsec:select} and is 
closer to a binary search using rank queries.

%\paragraph{Structures Tested.} 
We refer to the implementation of our data 
structure as BG, for {\em block graph}. We compared space and time performance 
of BG with other relevant data structures supporting rank, select, and access
operations. These included: GCC, a grammar-compressed structure by
Navarro and Ord\'o\~{n}ez~\cite{NO14}; CM, an efficient implementation of the classical 
succinct (but not compressed) solution by Clarke and Munro~\cite{Mun96}; and RRR, the 
widely used $H_0$-compressed data structure of Raman et al.~\cite{RRR07}. 

All test were run on an Intel(R) Xeon(R) E5620 at $2.40$GHz with $96$GB of RAM. 
The OS was Ubuntu 10.04 with kernel 2.6.32-33-server.x86\_64. All implementations 
were written in {\tt C++}. The compiler was \verb|g++| version $4.6.3$, with \verb|-O9| 
optimization.

For test data, we built suffix trees for two different collections: {\sf einstein},
a collection of Wikipedia files with full version history; and {\sf influenza} a 
collection of hundreds of individual genomes of influenza viruses. The suffix tree topologies
were represented as sequences of balanced parentheses (and are thus binary strings). For {\sf influenza}, the 
length of this string was 603,704,964 and parsed into 1,133,015 LZ factors. 
The string for {\sf einstein} was 367,324,468 bits long and parsed into just 50,541
LZ factors.

\begin{figure}[htb]
\minipage{0.49\textwidth}
  \includegraphics[width=\linewidth]{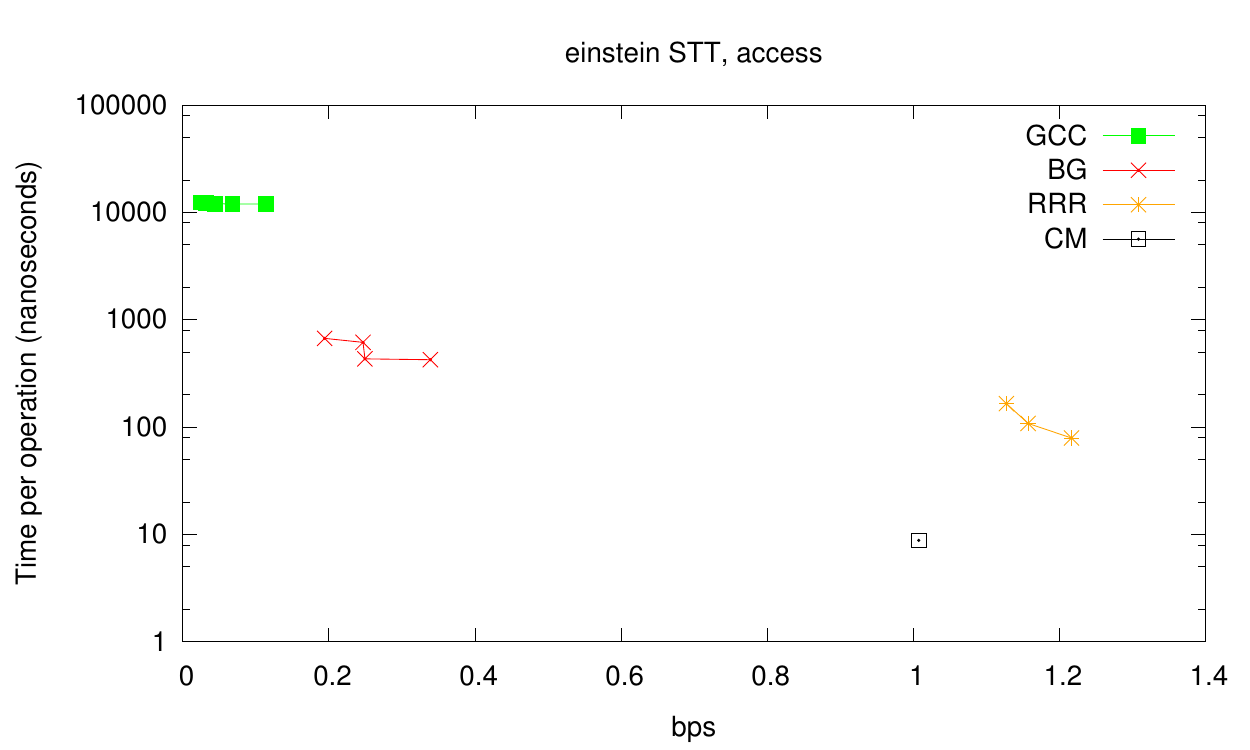}
\endminipage\hfill
\minipage{0.49\textwidth}
  \includegraphics[width=\linewidth]{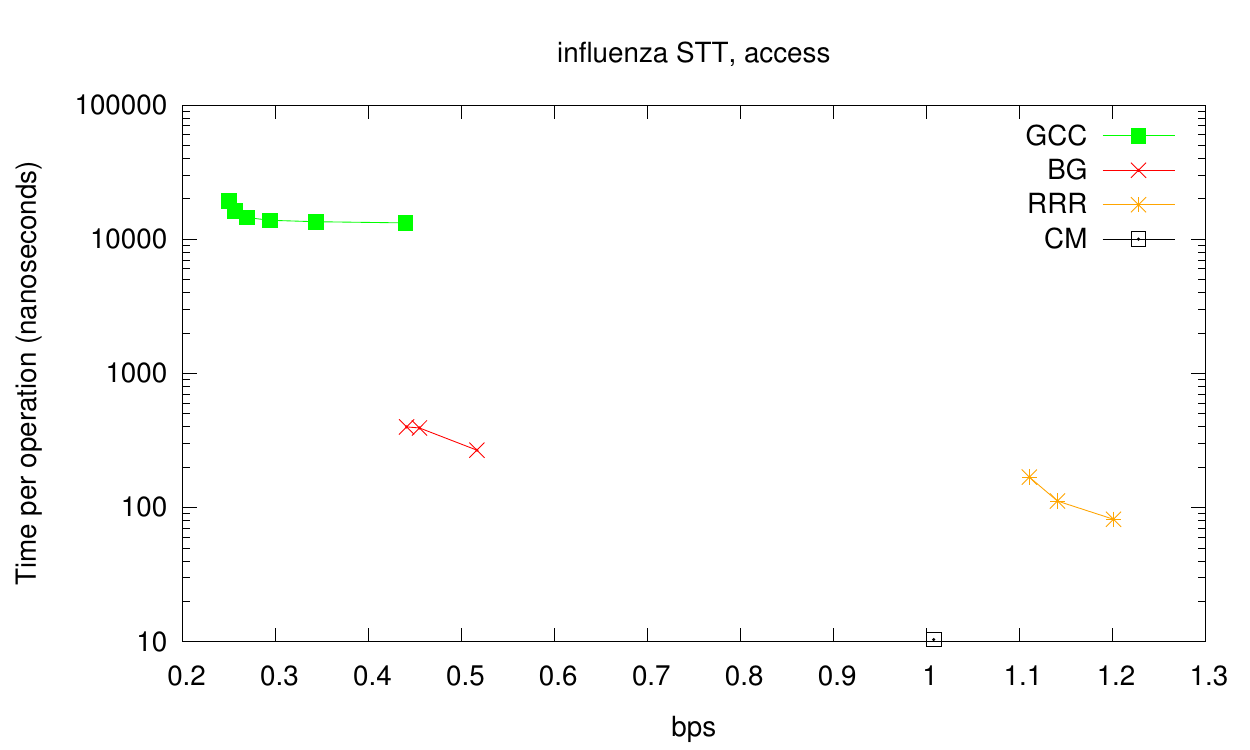}
\endminipage
\vspace{1ex}
\newline
\minipage{0.49\textwidth}
  \includegraphics[width=\linewidth]{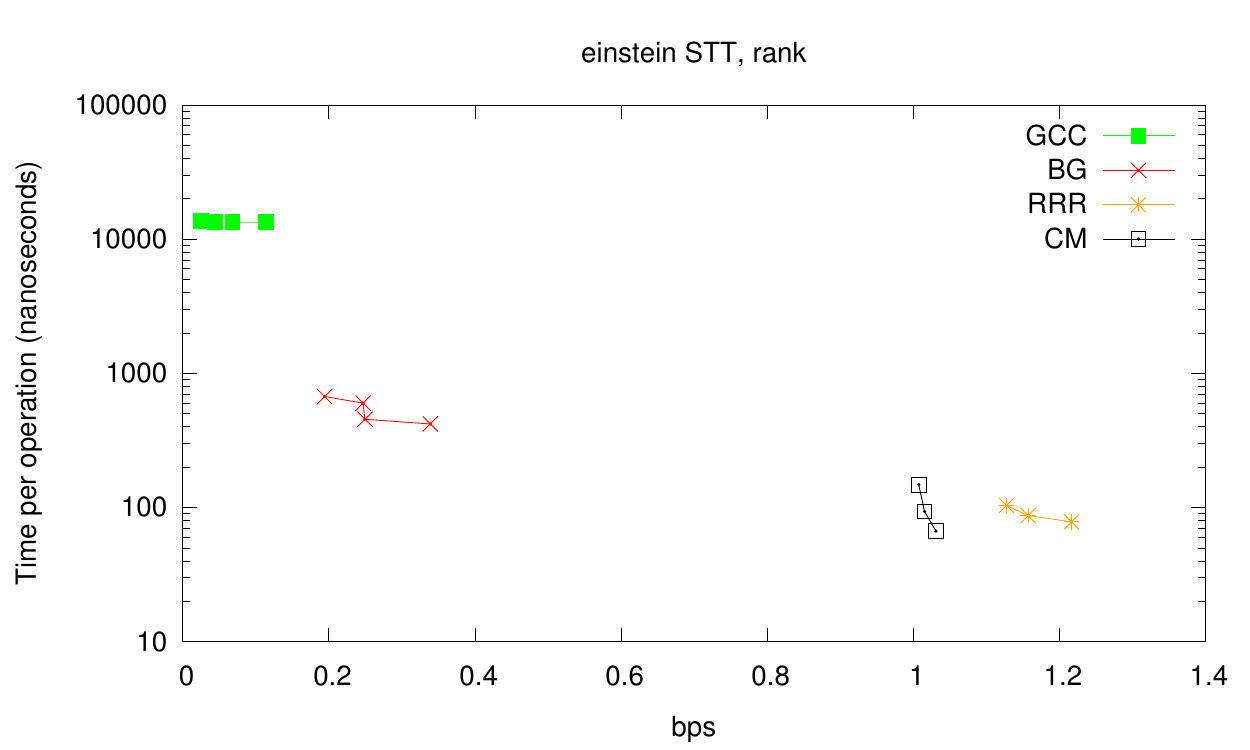}
\endminipage\hfill
\minipage{0.49\textwidth}
  \includegraphics[width=\linewidth]{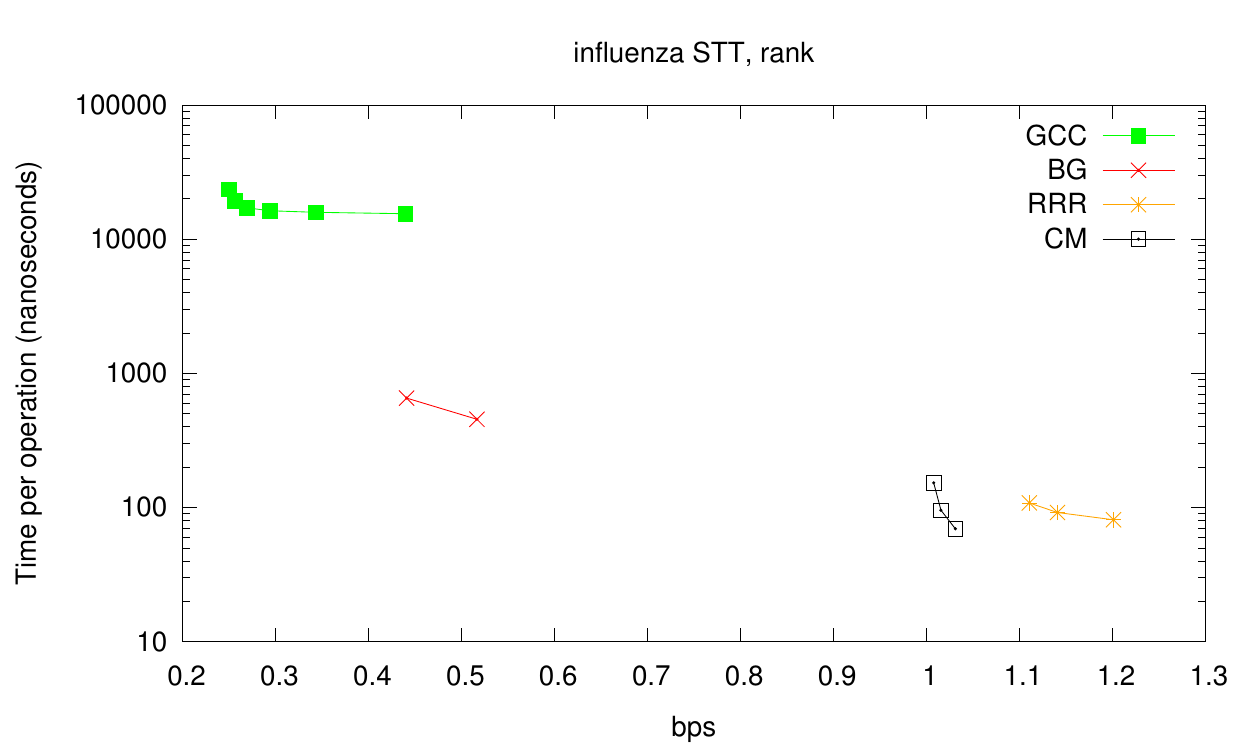}
\endminipage
\vspace{1ex}
\newline
\minipage{0.49\textwidth}
  \includegraphics[width=\linewidth]{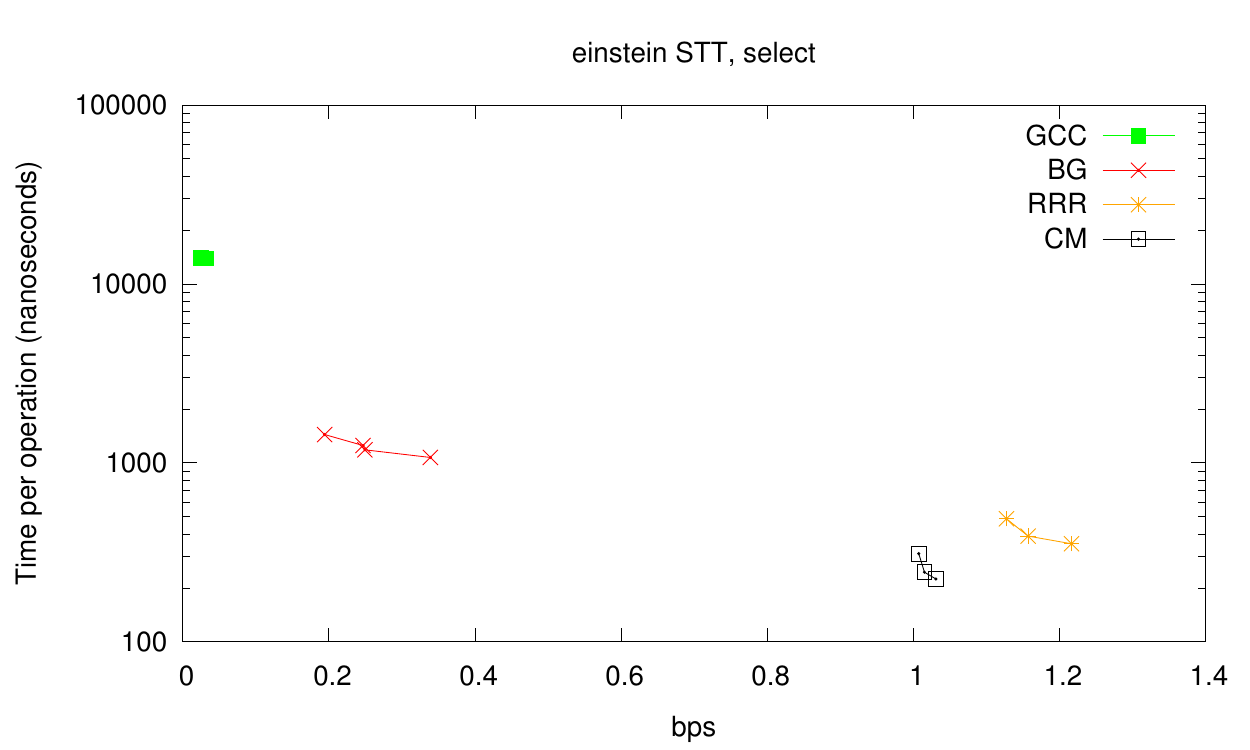}
\endminipage\hfill
\minipage{0.49\textwidth}
  \includegraphics[width=\linewidth]{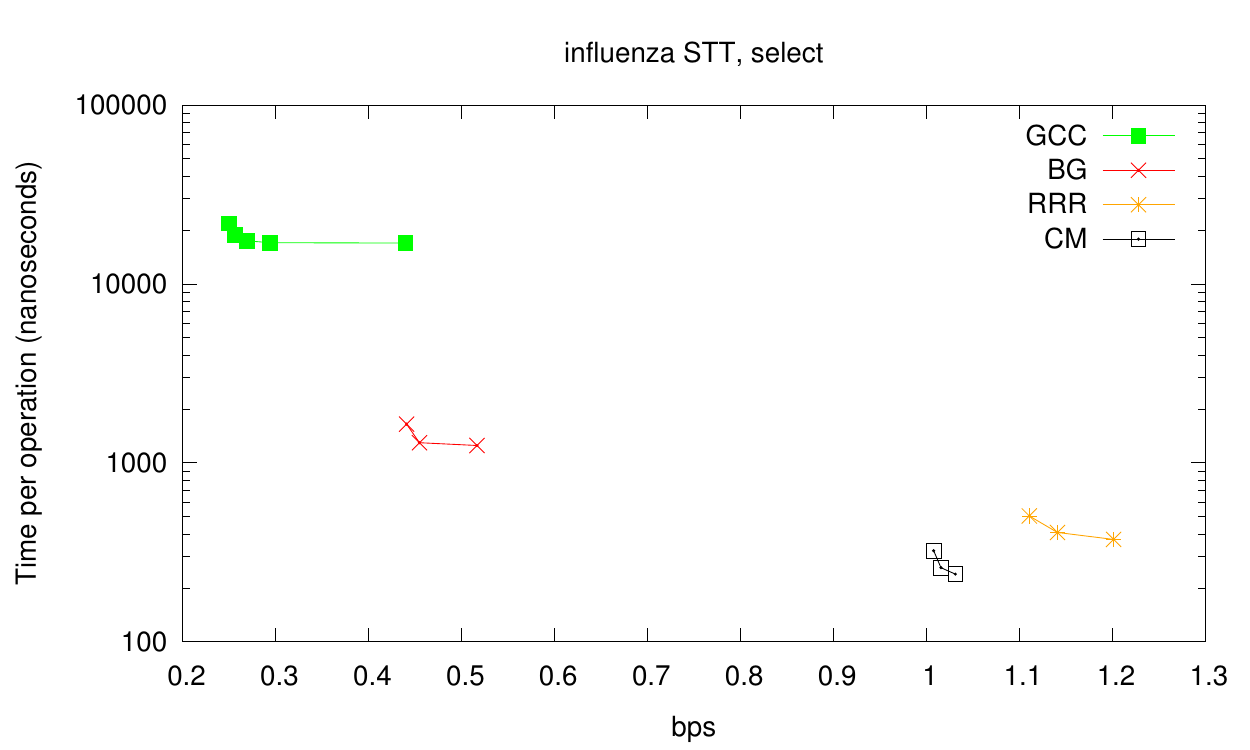}
\endminipage

\caption{Space-time trade-offs for access (top), rank (middle) and select (bottom) queries
of various methods when applied to two suffix tree topologies.
Note the log scale on the y-axes and that not all x-axes start at 0.
}
\label{figure:bitvectors}
\end{figure}

Results are shown in Figure~\ref{figure:bitvectors}. 
BG is more than an order of magnitude faster than the grammar compressed data structure GCC
on all operations. GCC is capable of greater compression, but in many applications the 
tradeoff achieved by the block graph is much more useful and spans the long gap between
GCC and CM. The trade-off is achieved via parameter $r$, the arity of the block graph. 

The uncompressed solution CM is consistently fastest, 
as expected. Notably, because the repetition in these binary strings is non-local, the 
RRR method is unable to achieve any compression and is actually larger than the 
uncompressed CM structure, due to the dominance of lower-order terms.

\end{document}